\documentstyle[12pt]{article} 
\begin{document}
\begin{center}
 {\bf Symmetry and collective motion} \\[2cm] 
M. Grigorescu \\[3cm]  
\end{center}
\noindent
$\underline{~~~~~~~~~~~~~~~~~~~~~~~~~~~~~~~~~~~~~~~~~~~~~~~~~~~~~~~~
~~~~~~~~~~~~~~~~~~~~~~~~~~~~~~~~~~~~~~~~~~~}$ \\[.3cm]
This work presents the basic elements of the formalism involved in
the treatment of Hamiltonian dynamical systems with symmetry
and the geometrical description of collective motion.
 \\
$\underline{~~~~~~~~~~~~~~~~~~~~~~~~~~~~~~~~~~~~~~~~~~~~~~~~~~~~~~~~
~~~~~~~~~~~~~~~~~~~~~~~~~~~~~~~~~~~~~~~~~~~}$ 
\newpage
\section{Introduction}
The continuous symmetries of physical systems are reflected not only by conservation laws, but also by 
constraints on the probability, or particle distribution functions, and phase-space reduction. \\ \indent 
Similarly to the cohomology groups \cite{fh}, defined in terms of finite coverings by open sets, the 
statistical distributions are defined with respect to the partition of the phase-space in 
"elementary cells". In general, this space is the cotangent fibration over a configuration 
space which is homogeneous for the 
action of a kinematical symmetry group, such as $SO(3, {\sf R})$ or Galilei. When a many-particle 
equilibrium distribution of minimum energy breaks to some extent this symmetry, then it defines an 
"intrinsic frame", and the low-energy dynamics of the system can be described in terms of collective
modes.   \\ \indent
This work presents\footnote{the next sections are based on the notes
of the seminars "Lie algebras" and "Dynamical symmetries and collective
variables" given in 1987-1989 at the Institute of Atomic Physics from
Bucharest.}, following \cite{am,gs} as main references, the elements of the formalism involved in 
the treatment of Hamiltonian dynamical systems with symmetry. The "momentum map" provided by the set of 
invariant observables is presented in Section 2. The related Marsden-Weinstein reduction of the phase-space 
is discussed in Section 3. The geometrical approach to collective motion is formulated and illustrated 
in examples in Section 4.    
\section{Symplectic actions and the moment map}
Let $M$ be a smooth manifold. An action to the left of the Lie group $G$ on $M$ is a smooth map $\Phi : G \times M \mapsto M$ such that \\ 
{\it  i)} $\forall m \in M, \Phi (e,m) =  m $, \\ 
{\it  ii)} $\forall g,h \in G$, $\forall m \in M$, $\Phi(g, \Phi(h,m))  =  \Phi(gh,m)$.  \\
The orbit of $m \in M$ is  
$$ G \cdot m = \{ \Phi (g, m) / g \in G \}~~. $$ 
The action is transitive if $G \cdot m =M, ~ \forall m \in M $. In this case $M$ is called a 
homogeneous space. The isotropy group of $\Phi$ at $m \in M$ is  
$$ G_m = \{ g \in G / \Phi (g, m) =m \}~~. $$
This subgroup is closed because $\tilde{\Phi}_m : G \mapsto M$, $\tilde{\Phi}_m (g) \equiv \Phi_(g,m)$ is continuous, and $G_m = \tilde{\Phi}_m^{-1} (m)$. \\ \indent
Let $\Phi$ be an action of $G$ on the symplectic manifold $(M, \omega)$. This action is 
symplectic if $\omega$ is invariant to $\Phi_g$ ($\Phi_g (m) \equiv \Phi (g, m)$),    
$$ \Phi_g^* \omega = \omega~~,~~ \forall g \in G~~. $$
Let $\Phi : G \times M \mapsto M$ be a smooth action, and $\xi \in T_e G \equiv {\sf g}$.  
Then $\Phi^\xi : {\sf R} \times M \mapsto M$ defined by
$$ \Phi^\xi (t,m) = \Phi (e^{t \xi},m)~~,~~t \in {\sf R} $$ 
is an ${\sf R}$-action on $M$, and $\Phi^\xi$ is a current on $M$. The corresponding vector field given by 
$$ \xi_M(m) = \frac{d}{dt} \vert_{t=0} \Phi (e^{t \xi},m)$$
is the infinitesimal generator of the action induced by $\xi$, and 
$$\xi_M(\Phi_gm) = T_m \Phi_g (Ad_{g^{-1}} \xi )_M (m)$$
with $Ad_g \xi = T_e R_{g^{-1}} L_g \xi$ (Appendix 1).  \\
{\it Proposition 1.} Let $\Phi$ be an action to the left of $G$ on $M$, and $\tilde{\Phi}_m : G \mapsto M$ defined by $\tilde{\Phi}_m(g)= \Phi(g,m)$, $\forall m \in M$. Then: \\
{\it i)} for any $ \omega \in \Omega^q(M)$, $\Phi_g$-invariant ($\Phi_g^* \omega = \omega$), one can define a 
map\footnote{ $\Lambda^k({\sf g})$ is the set of $k$-forms on ${\sf g}$ (the left-invariant 
subset of $\Omega^k(G)$), $Z^k({\sf g}) = \{ \omega \in \Lambda^k({\sf g}) , d \omega =0 \}$,
$H^k( {\sf g} ) = Z^k( {\sf g}) / B^k({\sf g})= Ker~ d (\Lambda^k) / d \Lambda^{k-1}$.}  
 $\Sigma : M \mapsto \Lambda^q({\sf g})$,
$$ \Sigma(m) = \tilde{\Phi}_m^* \omega ~~.$$ 
{\it ii)} the map $\Sigma$ is a $G$-morphism (equivariant): \\
$$ \Sigma \circ \Phi_g = Ad_{g^{-1}}^* \Sigma ~~.$$
{\it Theorem 1.} Any symplectic action of a Lie group $G$ on $(M, \omega)$ defines a 
G-morphism $\Sigma : M \mapsto Z^2({\sf g})$ such that: \\
{\it i)} $\Sigma (M)$ is a union of G-orbits in  $Z^2({\sf g})$. \\
{\it ii)} if the action of $G$ on $M$ is transitive, then $\Sigma (M)$ is a single orbit. \\
Let $\Sigma \in Z^2({\sf g})$, ${\sf g} \equiv T_e G$, considered as a 2-form on $G$, and
$${\sf g}_\Sigma = \{ \xi \in {\sf g} / L_\xi \Sigma =0 \}~~,~~h_\Sigma =   
\{ \xi \in {\sf g} / i_\xi \Sigma =0 \}~~,$$
where $L_\xi = i_\xi d + d i_\xi$ denotes the Lie derivative. 
Thus, ${\sf g}_\Sigma$ is the isotropy algebra of $\Sigma$, while the
connected group $H_\Sigma$ generated by ${\sf h}_\Sigma$ is the leaf through the
identity $e \in G$ of the characteristic (null) foliation of $\Sigma$. \\
{\it Proposition 2. } For $\Sigma \in Z^2({\sf g})$, if $H_\Sigma$ is closed in $G$, then $M_\Sigma = G / H_\Sigma $ is a symplectic manifold, and the orbit in $Z^2({\sf g})$ associated with $M_\Sigma$ is the orbit of $G$ through $\Sigma$. \\
{\it Proof:} Let $\pi : G \mapsto M_\Sigma $ be the projection $\pi (g) = g H_\Sigma $. Considering $H_\Sigma \equiv m \in M_\Sigma$, then $\pi_m (g) = g m = \tilde{\Phi}_m(g)$, 
and $\pi = \tilde{\Phi}_m$, $\Sigma = \pi^*_m \omega$ $\bigtriangledown$. \\
{\it Proposition 3. } The $G$-orbits through the elements $\Sigma \in Z^2({\sf g})$ having $H_\Sigma$ closed are covering spaces for symplectic manifolds $(M, \omega)$ on which $G$ acts transitively. \\
{\it Proof:} Let $M=G/H$. If $i_{\eta_M} \omega =0$ then $\eta_M =0$ and $\eta \in {\sf h}$. 
Thus, if $\Sigma = \pi^* \omega $ then ${\sf h}_\Sigma = \{ \xi \in {\sf h} \} ={\sf h}$, 
 $H_\Sigma$ is the connected component of $H$, and  $M_\Sigma=G/H_\Sigma$ is covering space for $M$ $\bigtriangledown$. \\
{\it Example.} (Kostant - Souriau): If $\Sigma = d \beta$, $\beta \in {\sf g}^*$, then 
$H_\Sigma$    is closed, and represents the connected component of the group
$$ H_\beta = \{ g \in G / Ad_{g^{-1}}^* \beta = \beta \}~~. $$      
Thus, $G/H_\Sigma$ is covering space for the orbit $G \cdot \beta \simeq G / H_\beta$. \\
{\it Theorem 2}, (Kostant-Souriau): If $H^1({\sf g})=H^2({\sf g}) = \{ 0 \}$, then any 
symplectic manifold, homogeneous for $G$, covers a $G$-orbit in ${\sf g}^*$ \cite{nrw}. \\
{\it Proof }. If $H^2({\sf g}) = \{ 0 \}$ then $\forall \omega \in Z^2({\sf g})$ has the form $\omega = d \beta$, $\beta \in {\sf g}^*$, and $H_\omega$ is closed. However,if  $H^1({\sf g}) = \{ 0 \}$ then $\forall \beta \in Z^1({\sf g})$, $\beta =0$, so that $\forall \omega \in Z^2({\sf g})$, 
$\omega = d \beta$ with $\beta$ unique $\bigtriangledown$. \\
{\it Proposition 4}. Let us consider $\omega \in Z^2({\sf g})$, left-invariant on $G$. Then: \\
{\it i)} the subalgebra ${\sf h}_\omega$ of minimal dimension is commutative. \\ 
{\it ii)} if $M$ is a homogeneous symplectic space with maximal dimension, then $\forall m \in M$, the connected component of the isotropy group $G_m$ is abelian. \\
{\it Proof}. {\it i)} Assume $\omega_t = \omega + t d \theta$ with $\theta \in \Omega^1(G)$, 
left-invariant. If $dim({\sf h}_\omega)$ is minimal, then $dim({\sf h}_{\omega_t})=
dim({\sf h}_\omega)$, and $\forall \xi \in {\sf h}_\omega$ can be extended to a curve 
$\xi_t = \xi + t \eta + 0(t^2) \in {\sf h}_{\omega_t}$. As $i_{\xi_t} \omega_t =0$, 
$\theta([ \xi, \eta]) =0$ $\forall \theta$, such that $[\xi, \eta]=0$. \\
{\it ii)} If $M \simeq G/H$ has maximal dimension then $H$ has minimal dimension, and 
${\sf h} = {\sf h}_\omega$ is abelian $\bigtriangledown$.   \\ \indent   
Let $\Phi: G \times M \mapsto M$ be a symplectic action of $G$ on $(M, \omega)$. The action $\Phi$ is 
strongly symplectic\footnote{ If $H^1( M, {\sf R}) = \{0 \}$ then every closed 
1-form is exact, and every symplectic action is strongly symplectic.} (Hamiltonian) if there 
exists a (comomentum) map 
$$\lambda : {\sf g} \mapsto {\cal F}(M)~~, $$
such that $i_{\xi_M} \omega  = d \lambda_\xi$. A strongly symplectic action $\Phi$ is 
strongly Hamiltonian if $\lambda$ is a $G$-morphism (equivariant), namely
$$ \Phi^*_{g^{-1}} \lambda = \lambda \circ Ad_g~~, $$
and the diagram 
\\ 
\begin{center} 
\( \begin{array}{ccc}
 0 ~ \mapsto ~{\sf R} ~  \mapsto & {\cal F} (M) ~ \mapsto  ~ ham(M)  &  \mapsto ~ 0 \\ 
~~~~~~~ & ~~~~~_{\lambda}  \nwarrow ~~  \uparrow_{~~~~~~} & \\ 
~~~~~~~ & ~~~~~~~~ {\sf g} &  
\end{array} \) 
\\ 
\end{center} 
commutes. \\ \indent
Let $\Phi: G \times M \mapsto M$ be a symplectic action on the connected symplectic manifold $(M, \omega)$. 
Then $J: M \mapsto {\sf g}^*$ is a moment map for the action $\Phi$ if $\forall \xi \in 
{\sf g}$ $$i_{\xi_M} \omega = d \hat{J} ( \xi) $$ where $\hat{J}(\xi) : M \mapsto {\sf R}$ is defined by 
$\hat{J}(\xi)_m = J_m \cdot \xi \equiv \langle J_m , \xi \rangle$. In particular, a Hamiltonian action $\Phi : G \times M \mapsto M$ defines 
a moment map by $\hat{J} (\xi) = \lambda_\xi$.  \\ 
{\it Proposition 5.} Let $M$ be a symplectic connected manifold and $\Phi :G \times M \mapsto M$ a 
symplectic action with the moment map $J$. For any $a \in G$ and $\xi \in {\sf g}$ one defines the map 
$\hat{Z}_a (\xi) :M \mapsto {\sf R}$,
$$\hat{Z}_a(\xi) = \Phi^*_{a^{-1}} \hat{J} (\xi) - \hat{J} (Ad_a \xi)~~. $$
Then \\
{\it i)}  $\hat{Z}_a(\xi)$ is a constant on $M$ and defines a map $Z : G \mapsto {\sf g}^*$,
$ \langle Z_a, \xi \rangle = \hat{Z}_a(\xi)$, called coadjoint cocycle. \\
{\it ii)} $Z$ satisfies the identity $Z_{ab} = Z_a + Ad^*_{a^{-1}} Z_b $. \\
{\it Proof.} $M$ is connected and $d\hat{Z}_a(\xi) =0$ on $M$, such that $\hat{Z}_a(\xi)$ is a constant $\bigtriangledown$. \\ \indent
One should note that the cocycle $Z : G \mapsto {\sf g}^*$ defines an infinitesimal cocycle $\omega \in Z^2({\sf g})$, by
\begin{equation}
\omega (\xi, \eta) = \frac{d}{dt} \vert_{t=0} \hat{Z}_{\exp (\xi t)} (\eta) = \{\hat{J}(\xi), \hat{J}(\eta) \} - \hat{J}([\xi, \eta])~~. \label{infc1}
\end{equation}
A coadjoint cocycle $\Delta : G \mapsto {\sf g}^*$ is coboundary if there exists $\mu \in {\sf g}^*$ such that $$\Delta_a = \mu - Ad_a^* \mu~~.$$
Similarly to (\ref{infc1}), $\Delta$ defines an infinitesimal coboundary $\delta \in B^2({\sf g})$ by
$$\delta (\xi, \eta) = \frac{d}{dt} \vert_{t=0} \hat{\Delta}_{\exp (\xi t)} (\eta) = -
\mu \cdot [\xi, \eta]~~,~~ \delta = d \mu~~.$$            
The set of the equivalence classes $[Z]$ of the cocycles-mod-coboundaries is the (deRham) 
cohomology space\footnote{ $b_k (M)= dim ~ H^K(M, {\sf R})$, $k=0,1,2,...$ are the Betti 
numbers, and $\chi(M, {\sf R}) = \sum_k  (-1)^k b_k(M)$ is the Euler-Poincar\'e 
characteristic of M.} $H^2(G,{\sf R})$ of $G$.  \\ 
{\it Proposition 6.} Let $J_1$ and $J_2$ be two moment maps for the symplectic action $\Phi : G \times M \mapsto M$, with cocycles $Z_1$ and $Z_2$. Then $[Z_1]=[Z_2]$, so that to any symplectic action which admits a moment map it corresponds a well defined cohomology class. 
\\ 
{\it Proposition 7.} Let $\Phi : G \times M \mapsto M$ be a symplectic action with a moment map. Then \\
{\it i)} $\Phi$ defines a cohomology class $[c_\Phi] \in H^2({\sf g},{\sf R})$ which 
measures the obstruction to find an equivariant moment map ($J \circ \Phi_g = Ad^*_{g^{-1}} J$). \\
{\it ii)} an equivariant moment map $J$ exists only iff $[c_\Phi]=0$. \\
{\it iii)} when $[c_\Phi]=0$ the set of all possible equivariant moment maps is parameterized by $H^1({\sf g})$. \\
{\it Examples} \\
I. If ${\sf g}$ is a semisimple Lie algebra (Appendix 2) then $H^1({\sf g})=H^2({\sf g})=0$, and any symplectic action of $G$ admits an unique equivariant moment map. \\
II. Let $G = H \times V$ be the Galilei group, expressed as a semidirect product between $H=SO(3,{\sf R}) \times {\sf R}$ and $V={\sf R}^6$. Thus, an element $\Gamma \in G$, 
\begin{equation} 
\Gamma = \left[ \begin{array}{ccc} \hat{A} & {\bf V} & {\bf X} \\ 
0 & 1 & t \\
0 & 0 & 1  \end{array} \right] ~~,
\end{equation}  
is specified by $\hat{A} \in SO(3,{\sf R})$, $t \in {\sf R}$, ${\bf V} \in {\sf R}^3$ and ${\bf X} \in {\sf R}^3$.  Such an element acts on a point $({\bf x}_0,t_0)$ of the space-time manifold $M={\sf R}^3 \times {\sf R}$ by
\begin{equation}
\Phi_\Gamma ({\bf x}_0,t_0) = (\hat{A} {\bf x}_0 + {\bf X} + {\bf V} t_0, t+t_0)~~.
\end{equation}
The algebra ${\sf g}$ of $G$ has the form ${\sf g}={\sf k}+{\sf p}$, where ${\sf k} = \{ \hat{\xi} \in so(3) \}$, ${\sf p} = \{  ({\bf v}, {\bf x}, \tau) \in {\sf R}^7 \}$, ${\sf k} \cap {\sf p} = \{ 0 \}$, $[{\sf k},{\sf k}] \subset {\sf k}$, $[{\sf k},{\sf p}] \subset {\sf p}$ according to 
$$ [ \hat{\xi} , ({\bf v}, {\bf x}, \tau) ] = ( \hat{\xi} {\bf v}, \hat{\xi} {\bf x}, 0)~~, $$ and $[{\sf p},{\sf p}] \subset {\sf p}$,
$$ [ ({\bf v}, {\bf x}, \tau) , ({\bf v}', 
{\bf x}', \tau') ] = ( 0, \tau' {\bf x} - \tau {\bf v}', 0)~~. $$
{\it Proposition 8.} If $G$ is the Galilei group\footnote{The result for the Lorentz group is presented in
\cite{et}.}, then $H^2({\sf g})$ is 1-dimensional, and up to a coboundary any cocycle $\Sigma \in Z^2({\sf g})$ is given by
\begin{equation}
\Sigma((\hat{\xi}, {\bf v}, {\bf x}, \tau),(\hat{\xi}',{\bf v}', {\bf x}', \tau')) =
m({\bf v} \cdot {\bf x}' - {\bf v}' \cdot {\bf x}) ~~.
\end{equation} 
{\it Theorem 3.} Let $\Phi$ be a transitive Hamiltonian action of $G$ on $M$ with the 
equivariant moment map $J$. Then: \\
{\it i)} $J$ maps $M$ on an orbit of $G$ in ${\sf g}^*$. \\
{\it ii)} $J$ is an immersion, and $M$ is a covering space for $J(M)$. \\
{\it Proof.} {\it (i)} $ \Phi^*_{g^{-1}} J = J \circ Ad_g$, with $\Phi_g$ transitive. 
{\it (ii)} Let $T_m J \cdot \eta_M =0$, $\eta \in {\sf g}$, $\eta_M \in T_mM$. Then 
$$ \langle T_m J \cdot \eta_M, \xi \rangle =0 ~~,~~ \forall \xi \in {\sf g}~~.$$  
If $dJ_\xi \cdot \eta_M=0$ then $\omega (\xi_M, \eta_M) =0$ $\forall \xi \in {\sf g}$. 
However $\{ \xi_M(m) / \xi \in {\sf g} \} = T_mM$ and $\omega$ is nondegenerate, such that 
$\eta_m =0$. Thus, ${\rm Ker} (TJ) = \{ 0 \}$, and $J$ is an immersion $\bigtriangledown$.
\section{Dynamical systems with symmetry} 
\subsection{ Introduction} 
The description of space and time in classical mechanics in terms of a class of reference frames which are 
equivalent with respect to formulation of the principles of mechanics extends the 
notion of geometrical symmetry to the one of symmetry of equations of classical mechanics. This 
symmetry has a kinematical character, because is independent of the physical system under consideration, 
and the effect of symmetry transformations is a change in the reference frame. As the set of geometrical 
symmetry transformations is a group, the kinematical symmetry transformations form a Lie group, called 
Galilei group in the nonrelativistic case, or Poincar\'e in relativistic mechanics. The existence of these 
symmetry groups is reflected by the existence of a set of observables $\{ J_\alpha \}_{\alpha \in I}$ 
which remain invariant during the time-evolution of the physical system. The fields $\{ X_\alpha \}_{\alpha 
\in I}$ on the classical phase-space $(M, \omega )$, $i_{X_\alpha} \omega = dJ_\alpha$,  generated by  
$\{ J_\alpha \}_{\alpha \in I}$, commute with the field $X_H$, $i_{X_H} \omega = dH$ of the Hamilton 
function, and $H$ is a constant on the leaves $J^{-1}_\alpha ( \mu_\alpha) \subset M$ determined by regular 
values  $\{ \mu_\alpha \}_{\alpha \in I}$. As a result, the "intrinsic" dynamics provided by $H$ is defined 
on a phase-space $\bar{M}$, called "reduced phase-space", with a dimension smaller than of the initial 
phase-space $M$. \\ \indent
The existence of additional invariant observables, beside the ones
provided by the kinematical symmetry, implies the existence of a larger
symmetry group, specific for the physical system under consideration, called
dynamical symmetry group (in the case of nonrelativistic Kepler problem this is $SO(3,1)$ 
for states with positive energy, $H=E>0$, and $SO(4,{\sf R})$ for bound states, $H=E<0$). Its
existence implies a lower dimensionality of the reduced phase-space, and
allows a partial description of the intrinsic dynamics in terms of the group
action. When the reduced phase-space $\bar{M}$ with respect to the action of
the dynamical symmetry group $G$ is a point, the system is called completely
integrable, and the Hamiltonian current on $M$ is provided by the action of
$G$, or one of its subgroups. In general, a similar result is obtained for an
equilibrium point on $\bar{M}$ of the reduced Hamiltonian, or for systems
with constraits which "freeze" the dynamics on $\bar{M}$. In such a situation
the system has a collective behaviour, because the Hamiltonian current on $M$
arises by the action of $G$, and can be described completely using a number
of variables smaller than the dimension of $M$.                   
\subsection{The reduced phase-space}
Let $(M, \omega)$ be a symplectic manifold, $\Phi : G \times M \mapsto M$ a symplectic 
action of $G$ on $M$, $J:M \mapsto {\sf g}^*$ a moment map for $\Phi$ such that  
$ J \circ \Phi_g = Ad_{g^{-1}}^* J$, and $$ G_\mu = \{ g \in G /  Ad_{g^{-1}}^* \mu = \mu \}~~,$$ with 
$\mu$ a regular value\footnote{For $f:M \mapsto N$ of class $C^1$, $n \in N$ is a regular value of $f$ if $\forall m \in f^{-1}(\{ n \}) $, $T_mf$ is surjective.}
  of $J$. Then: \\
{\it i)} $P_\mu = J^{-1}(\mu)$ is a submanifold of $M$. \\
{\it ii)} $\forall p \in J^{-1}(\mu)$, $T_p (G_\mu \cdot p) = T_p (G \cdot p) \cap T_p P_\mu$, such that $G_\mu$ acts on $J^{-1}(\mu)$. \\
The proof of {\it (i)} follows from the Sard's theorem, while for {\it (ii)}   
$ J \circ \Phi_g = Ad_{g^{-1}}^* J$ shows that $\xi_M(p) \in T_p J^{-1}(\mu) $ iff $\xi_{{\sf g}^*}(\mu)=0$, namely $\xi \in {\sf g}_\mu$ $\bigtriangledown$. \\ \indent
An action $\Phi : G \times M \mapsto M$ is called proper if the map $\Psi    
: G \times M \mapsto M \times M$ defined by $\Psi(g,m) = (m, \Phi_g (m)) $ is proper\footnote{ Which means that 
any compact subset $K \subset M \times M$ is the image of a compact subset 
$\Psi^{-1}(K) \subset G \times M$, or that if the sequences $x_i$  and $\Phi_{g_i} x_i$, 
$i=1,2,3,...$, are convergent, then $g_i$ contains a convergent subsequence.}. For such an action $\forall m \in M$, 
$G_m = \{ a \in G / \Phi_a(m)=m \}$ is compact. The action is called free if $G_m = \{ e \}$. \\ \indent
If $\mu$ is a regular value of $J$, then the action of $G_\mu$ on $J^{-1}(\mu)$ is locally free, and provides a foliation having as leaves the orbits of $G_\mu$. If the action of $G_\mu$ is proper, then the orbits $G_\mu \cdot p$, $p \in P_\mu$, are closed submanifolds in $J^{-1}(\mu)$. If the action of $G_\mu$ is free,     
$\forall p \in P_\mu$, $G_p = \{ e \} \subset G_\mu$, and there exists a submanifold $S \subset P_\mu$, containing $p$, having the properties: \\
{\it i)} $S$ is closed in $G_\mu(S)$. \\
{\it ii)} $G_\mu(S)$ is an open neighborhood of the orbit $G_\mu \cdot p$. \\
{\it iii)} if $a \in G_\mu$ and $\Phi_a (S) \cap S \ne \emptyset$ then $a = e$. \\
This last property shows that the leaves of $G_\mu$ through the points of $S$ may intersect at most once, so that the space of the orbits $\bar{M} = J^{-1}(\mu) / G_\mu$ has the structure of a manifold (the coordinates on $S$ can be choosen as local coordinates on $M$)\footnote{If the action of $G$ on $M$ is smooth, 
free and proper then $M \mapsto M/G$ is a principal bundle with $G$ as fiber and structure group,
\cite{am} p. 276.}. Moreover, $\bar{M}$ is Hausdorff because the orbits are closed submanifolds in $J^{-1}(\mu)$ ($G_\mu$ is closed and $\Phi$ is continuous). When {\it (iii)} does not hold $\forall S$, then $G_p \ne \{ e \}$. \\ \indent
Let $P$ be a manifold and $\omega \in Z^2(P)$. Then 
$$ E_\omega = \{ v \in TP / i_v \omega =0 \}$$ is called the characteristic distribution of $\omega$. 
If $E_\omega$ is a subbundle of $TP$, then $\omega$ is called regular on $P$. In such a case, the distribution $E_\omega$ is integrable, and it defines a foliation ${\cal F}$ on $P$. Let $\bar{M} = P/ {\cal F}$ be the space of leaves, obtained by the identification of the points from each leaf. The space $\bar{M}$ has the structure of a manifold if any point of a leaf is contained in a submanifold $S$, transversal to the leaf, such that $S$ intersects the leaf at most once. In this case, the local coordinates on $S$ can be used as local 
coordinates on $\bar{M}$, and $\bar{M}$ aquires the structure of a manifold.   \\ 
{\it Proposition 9.} Let $G$ be a compact Lie group acting on the manifold $P$ such that its orbits provide a foliation of $P$. For $p \in P$, $G_p = \{ a \in G / \Phi_a (p)=p \}$,
and $G_p^0$ denotes the connected component of $G_p$. Then the foliation of $P$ by $G$-orbits is fibration iff the representation of $\Gamma_p = G_p /G_p^0$ on the space $N_p$, 
normal to the orbit $G \cdot p$, is trivial $\forall p \in P$.   \\
{\it Examples of one-dimensional foliations which are not fibrations} \\
1. {\it The linear current on the 2-dimensional torus} \\
Let $T={\sf R}^2/{\sf Z}^2$ be the 2-dimensional torus, with coordinates 
$$[x,y]=(x,y) {\rm mod} {\sf Z}^2~~,~~(x,y) \in {\sf R}^2~~.$$
An action $\Phi_t:T \mapsto T$, $t \in {\sf R}$, of ${\sf R}$ on $T$, can be defined by
$$\Phi_t [x,y] =[x+at,y+bt]~~,$$ 
where $a,b$ are real constants. Thus, $\{ \Phi_t [x,y] / t \in {\sf R} \}$ is the leaf at $[x,y]$. When the ratio $a/b$ is irrational, any such orbit is dense on the torus. \\ \indent
Let $\bar{M}$ be the space of the leaves of the foliation, and $\pi :T \mapsto \bar{M}$,
$$ \pi [x,y] =  \{ \Phi_t [x,y] / t \in {\sf R} \}~~,$$ 
the projection on $\bar{M}$. Presuming that there exists a topology on $\bar{M}$ with respect to which $\pi$ is continuous, let $U$ be an open set on $\bar{M}$. If $U \ne \emptyset$, then $\pi^{-1} (U) =T$, so that $U=\bar{M}$. Thus, the only admissible topology on $\bar{M}$ is the trivial one. \\
2. {\it The $S^1$ orbits on the M\"obius band} \\
Let $\varphi : {\sf Z} \times {\sf R}^2 \mapsto {\sf R}^2$ be the action of ${\sf Z}$ on ${\sf R}^2$ given by 
$$ \varphi_n (x,y) = A^n (x,y)~~,n \in {\sf Z}, ~~A(x,y)=(x+1,-y)~~.$$
The M\"obius band is then $M={\sf R}^2/{\sf Z}$, with points 
$$[x,y]= \{ \varphi_n (x,y) / n \in {\sf Z} \} \in M~~.$$        
Also, $M$ is a bundle over the circle $S^1$, with projection $\pi :M \mapsto S^1$,
$$\pi[x,y] = x {\rm mod} {\sf Z}~~.$$
Let $\Phi: G \times M \mapsto M$ be the action on $M$ of the circle $G= {\sf R} / 2 {\sf Z}$, given by
$$ \Phi_t[x,y] =[x+t,y]~~,~~t \in {\sf R}~~.$$
The orbits of $G$ on $M$ are double coverings for the central circle $S^1 = \Phi_{\sf R} [x,0]$, and $\bar{M}$ is a half-line. It can be parameterized by $y \ge 0$, 
but is not a manifold near $y=0$. \\
{\it Theorem 4.} Let $(M, \omega)$ be a symplectic manifold, $G$ a Lie group acting symplectically on $M$, and 
$J : M \mapsto {\sf g}^*$ the equivariant moment map. Let $\mu \in {\sf g}^*$ be a regular value of $J$, and 
presume that $G_\mu$ acts freely and properly on $J^{-1}( \mu)  \equiv P_\mu$. If ${\rm i}_\mu : J^{-1}(\mu) 
\mapsto M$ is the inclusion, and ${\rm i}^*_\mu \omega = \omega \vert_{P_\mu}$, then: \\
The leaves of the characteristic distribution of the form ${\rm i}^*_\mu \omega$ on $P_\mu$ are orbits for $G_\mu$, and the orbit space $\bar{M} = J^{-1}(\mu) / G_\mu$ has the structure of a symplectic manifold $(\bar{M}, \bar{\omega})$, $\pi^* \bar{\omega} = {\rm i}^*_\mu \omega$, with $\pi : P_\mu \mapsto \bar{M}$ the projection $\pi (p) = \{ G_\mu \cdot p \}$. \\
{\it Proof.} $\forall \xi \in {\sf g}_\mu$, $i_{\xi_M} \omega \vert_{P_\mu} =0$, and 
$\xi_M \in E_{i^*_\mu \omega}$. Conversly, let $v$ be an element of $ T_p P_\mu$, but not 
of $T_p (G_\mu \cdot p)$, such that $i_v \omega \vert_{P_\mu}=0$. If $F= T_p (G \cdot p) 
\cup T_p P_\mu$ , then $F \cap F^{\perp} = T_p (G_\mu \cdot p)$. As $\omega$ is nondegenerate on $T_p M$, $F / F \cap F^{\perp}$ should be symplectic, with $\hat{\omega} ([f],[f']) = \omega (f,f')$, $f,f' \in F$. Thus, if $i_v \omega =0$, $v \in T_p P_\mu$, then also $i_{[v]} \hat{\omega} =0$, so that $[v]=0$ and $v \in T_p (G_\mu \cdot p)$ $\bigtriangledown$.\\
{\it Examples} \\
1. Let $f_1,f_2,...,f_n$ be $n$ functions in involution, $\{ f_i, f_j \} =0$, with respect to the Poisson backet on a $2N$-dimensional manifold $M$. Considering $G = {\sf R}^n$, $\mu \in {\sf R}^n$ and $J = f_1 \times f_2 \times ... \times f_n$, then 
$G_\mu=G$ and ${\rm dim} (J^{-1}( \mu) / G) = 2N -2n$. When $n=N={\rm dim} M /2$ the system is called completely integrable. \\
2. Let ${\cal H}$ be a complex Hilbert space with the symplectic form $\omega (X, Y) = Im \langle X \vert Y \rangle $, and $\Phi_z : {\cal H} \mapsto {\cal H}$, $\Phi_z  \psi = z \psi$, the action of $G =S^1 = \{ z \in {\sf C} / \vert z \vert =1 \}$ on ${\cal H}$. Then \\
{\it i)} the action $\Phi$ is symplectic. \\
{\it ii)} the moment map is $J_\psi = - i \langle \psi \vert \psi \rangle /2$. \\
{\it iii)} the reduced phase-space is the projective Hilbert space ${\sf P}_{\cal H}$. \\
{\it Proof.} The algebra of $G$ is ${\sf g} = \{ \xi \in {\sf C} / \xi = - \xi^* \}$, and as $\Phi_{e^{\xi t}} \psi = e^{\xi t} \psi$, $\psi \in {\cal H}$, $\xi_{\cal H} (\psi) = \xi \psi$. Also 
$$ T_\psi \Phi_z X_\psi = \frac{d}{dt} \vert_{t=0} \Phi_z ( \psi + X_\psi t) = 
 \frac{d}{dt} \vert_{t=0} z (\psi+X_\psi t) = z X_\psi$$
{\it i)} 
$\Phi_z^* \omega(X_\psi,Y_\psi) = \omega (T_\psi \Phi_z X_\psi, T_\psi \Phi_z Y_\psi) = 
\omega (zX_\psi,zY_\psi) = \omega (X_\psi,Y_\psi) $, or $\Phi_z^* \omega = \omega$. \\
{\it ii)} 
$$ i_{\xi_{\cal H}} \omega ( X_\psi) = Im \langle \xi_{\cal H}(\psi) \vert X_\psi \rangle =
Im \langle \xi \psi \vert X_\psi \rangle = i \xi Re \langle \psi \vert X_\psi \rangle \equiv 
\frac{d}{dt} \vert_{t=0} \hat{J} ( \xi)_{\psi+tX_\psi}$$
However, for $f(\psi) = \langle \psi \vert \psi \rangle$,
$$\frac{d}{dt} \vert_{t=0} f(\psi+tX) = 2 Re \langle \psi \vert X \rangle $$
such that $\hat{J} (\xi)_\psi = i \xi \langle \psi \vert \psi \rangle /2$. \\
{\it iii)} As ${\sf g}^* \simeq {\sf g}$, if $\mu \in {\sf g}^*$ and $\xi \in {\sf g}$ then 
$\langle \mu, \xi \rangle = \mu^* \xi$, and $J_\psi = - i \langle \psi \vert \psi \rangle /2$. Thus, 
$$J^{-1} ( \mu) = \{ \psi \in {\cal H} /  \langle \psi \vert \psi \rangle = 2 i \mu \}~~.$$
Because $G_\mu = G =S^1$ acts on $J^{-1} (\mu)$,
$$J^{-1}(\mu) /S^1 \simeq \{ [\psi] , \psi \in {\cal H} / \langle \psi \vert \psi 
\rangle =1 \} \equiv {\sf P}_{\cal H}~,~[\psi]= \{ z\psi / z 
\in {\sf C}, \vert z \vert =1 \}~~ \bigtriangledown .  $$ 
{\it Theorem 5.} With the assumptions of Theorem 4, let $(M, \omega)$ be a symplectic manifold, $G$ a Lie group 
acting symplectically on $M$, $J : M \mapsto {\sf g}^*$ the equivariant moment map, and $H : M \mapsto {\sf R}$ the 
Hamilton function, invariant to the action of $G$ (Appendix 3). If $\mu \in {\sf g}^*$ is 
a regular value of $J$, $G_\mu$ acts freely and properly on $J^{-1}( \mu)  \equiv P_\mu$, ${\rm i}_\mu : J^{-1}(\mu) \mapsto M$ is 
the inclusion, and ${\rm i}^*_\mu \omega = \omega \vert_{P_\mu}$, then: \\
{\it i)} the current $F_t$ of $X_H$ leaves $J^{-1}( \mu)$ invariant and commutes with the 
action of $G_\mu$ on $J^{-1}( \mu)$, such that it induces canonically a current $\phi_t$ on $\bar{M}$,
\begin{equation}
\pi_\mu \circ F_t = \phi_t \circ \pi_\mu ~~. \label{rc}
\end{equation} 
{\it ii)}  the current $\phi_t$ on $\bar{M}$ is Hamiltonian, generated by the function $\bar{H} : \bar{M} \mapsto {\sf R}$,
$$ \bar{H} \circ \pi_\mu = H \circ {\rm i}_\mu $$ 
called reduced Hamiltonian. \\
{\it Proof.} {\it (i)} As $J^{-1} (\mu)$ is invariant to $F_t$, the projection of $F_t$ on $\bar{M}$ defines the current $\phi_t$ by (\ref{rc}), and
$$ (\phi_t \circ \pi_\mu )^* \bar{\omega} = F_t^* \pi_\mu^* \bar{\omega} =  
F_t^* i_\mu^* \omega = i_\mu^* \omega = \pi_\mu^* \bar{\omega}~~.$$
Thus,  $\pi_\mu^* ( \phi_t \bar{\omega} - \bar{\omega} )=0$, and as $\pi_\mu$ is surjective, 
$\phi_t \bar{\omega} = \bar{\omega}$. \\
{\it (ii)}  Let $v \in TM$ and $[v] = T \pi_\mu v \in T \bar{M}$. Then 
$$ d \bar{H} ([v]) = \pi^*_\mu d \bar{H} (v) = d(\bar{H} \circ \pi_\mu) (v) =  
d(H \circ i_\mu) (v) = i_\mu^* dH (v) = i_\mu^* \omega (X_H, v) =$$
$$ \pi_\mu^* \bar{\omega} (X_H, v) = \bar{\omega} (T \pi_\mu X_H, T \pi_\mu v) = \bar{\omega}([X_H], [v]) = i_{[X_H]} \bar{\omega} ([v])~~, $$ 
such that $i_{[X_H]} \bar{\omega} = d \bar{H}$. Thus, $\phi_t$ has $[X_H]= T \pi_\mu X_H$ as 
generator and $\bar{H}$ as Hamiltonian $\bigtriangledown$. \\ 
{\it Proposition 10.} Let $c_t$, with $c_0 = m_0 \in J^{-1} (\mu)$, be an integral curve 
for $X_H$, and $[c_t]$ the integral curve for $X_{\bar{H}}$, presumed to be known. Let 
$d_t \in J^{-1} (\mu)$, with $d_0 =m_0$, a smooth curve such that $[d_t]=[c_t]$. Then $c_t$
is of the form
$$ c_t = \Phi_{a_t} d_t~~,$$
where $a_t \in G_\mu$ is provided by the equation $ \dot{a}_t = T_e L_{a_t} \xi (t)$
in which $\xi (t) \in {\sf g}_\mu$ is the solution of 
$$ \xi_M (d_t) = X_H (d_t) - \dot{d}_t~~.$$
{\it Proof.} If $c_t = \Phi_{a_t} d_t $ then
\begin{equation} 
X_H (c_t) = \dot{c}_t = T_{d_t} \Phi_{a_t} \dot{d}_t + T_{a_t} \Psi_{d_t} (a_t) 
\dot{a}_t~~, \label{10a}
\end{equation}
with $\Psi_m : G \mapsto M$, $\Psi_m (a) = \Phi_a m$. As 
$$ T_a \Psi_m \dot{a} = T \Psi_m T R_a T R_{a^{-1}} \dot{a} = T_e (\Psi_m \circ R_a)         
T R_{a^{-1}} \dot{a} = $$
$$T_e \Psi_{\Phi_a m} T R_{a^{-1}} \dot{a} = (T R_{a^{-1}} \dot{a})_M (\Phi_am)~~, $$
(\ref{10a}) becomes
\begin{equation}
X_H ( \Phi_a d_t) = T_{d_t} \Phi_a \dot{d}_t + (T R_{a^{-1}} \dot{a})_M (\Phi_a d_t)~~. 
\label{10b} 
\end{equation}
However, $X_H ( \Phi_a d_t)=T_{d_t} \Phi_a X_H(d_t)$ from the $\Phi_a$-invariance of $X_H$, 
and 
$$\xi_M ( \Phi_a d_t)=T_{d_t} \Phi_a (Ad_{a^{-1}} \xi)_M (d_t)~~,~~ Ad_a \xi \equiv T_e R_{a^{-1}} L_a \xi~~,$$ 
such that (\ref{10b}) takes the form 
$$X_H (d_t) = \dot{d}_t + (T_a L_{a^{-1}} \dot{a})_M (d_t)~~, $$
or $\xi_M (d_t) = X_H (d_t) - \dot{d}_t$ with $\xi =  T_a L_{a^{-1}} \dot{a}$ $\bigtriangledown$. \\
\subsection{Stability}
A point $m \in M$ is a relative equilibrium if $\pi_\mu (m) \in \bar{M}$ is a fixed point 
for the Hamiltonian system $X_{\bar{H}}$ on $\bar{M}$. If $\pi_\mu (m)$ is on a periodic orbit for 
$X_{\bar{H}}$ then $m \in M$ is called relative periodic. \\
{\it Proposition 11.} Let $m_0 \in J^{-1} (\mu)$ be a point of relative equilibrium for the Hamiltonian system $X_H$ on $M$, $X_H = \dot{F}_t \in \chi (M)$. Then there exists a one-parameter subgroup $a_t$ of $G_\mu$ such that
$$ m_t=F_t (m_0) = \Phi_{a_t} ( m_0)~~,~~ a_t \in G_\mu~~,$$
and $X_H (m_t) = \xi_M (m_t)$, $\xi \in {\sf g}_\mu$. \\ \indent
It is important to remark that if $m \in M$ is a relative equilibrium, or relative periodic 
of small amplitude, then the Hamiltonian system $X_H$ on $M$ has a collective behaviour\footnote{
If $\omega = - d \theta$ and $\Phi_a^* \theta = \theta$ then $L_{\xi_M} \theta =
i_{\xi_M} d \theta + d i_{\xi_M} \theta =0$, and $i_{\xi_M} \omega =  d \hat{J}( \xi)$ with
$\hat{J} (\xi)  = \theta (\xi_m)$.}. \\ \indent
Let $H$ be a Hamilton function on $(M, \omega)$, and $G$ a Lie group acting on $M$ with the moment map $J : M \mapsto {\sf g}^*$, so that $\Phi_a^* H =H$, $a \in G$. One defines the energy-moment map $H \times J :M \mapsto {\sf R} \times {\sf g}^*$ by
$$(H \times J) (m) = (H(m), J(m))~~.$$
The previous results indicate that $\forall {\cal C} \in {\sf R} \times {\sf g}^*$, ${\cal C} \equiv (E, \mu)$, the set $I_{\cal C} = (H \times J)^{-1} ({\cal C})$ is invariant to the curent of $X_H$. The topological structure of the current is determined by: \\
- the topological type of $I_{\cal C}$, $\forall {\cal C} \in {\sf R} \times {\sf g}^*$. \\
- the bifurcation set for the map $H \times J$. \\
- the current of $X_H$ on each $I_{\cal C}$. \\
- the decomposition of $H^{-1} (E)$ in submanifolds $I_{\cal C}$. \\
The knowledge of the bifurcation set and of the current of $X_H$ on each $I_{\cal C}$      may provide the values of ${\cal C}$ for which the system has a collective dynamics. 
\section{Collective dynamics} 
A classical collective model for the Lie group $G$ is a phase-space $(M, \omega)$ on which $G$ acts symplectically and transitively ($(M, \omega)$ is $G$-elementary). Let $\Phi : G \times M \mapsto M$ be a symplectic action of $G$ on $(M, \omega)$ with the moment 
map $J : M \mapsto {\sf g}^*$. A Hamiltonian $H$ on $M$ is called collective if it has the form
\begin{equation}
 H = h_c \circ J = J^* h_c~~, \label{hc}
\end{equation}
where $h_c : {\sf g}^* \mapsto {\sf R}$ is a smooth function. If $J(M)$ is a cotangent fibration, $J(M)= T^* Q$, 
then  (\ref{hc}) defines a physical collective model. \\ 
{\it Proposition 12.} Let $\Phi : G \times M \mapsto M$ be a Hamiltonian action of the Lie 
group $G$ on $(M, \omega)$ with the moment map $J :M \mapsto {\sf g}^*$, and 
$H = h_c \circ J$, $h_c: {\sf g}^* \mapsto {\sf R}$, a collective Hamiltonian. Then the 
trajectory $m_t$ of the Hamiltonian system $\dot{m}_t = X_H(m_t)$, $m_0 \in M$, is 
completely determined by the trajectory $\gamma_t$ of the Hamiltonian system defined by $h_c$ on the orbit ${\cal O} = G \cdot J(m_0)$, $\gamma_t = J(m_t)$. \\ 
{\it Proof.} $\forall \mu \in {\sf g}^*$, there exists ${\rm L}_{h_c} (\mu) \in {\sf g}$, 
${\rm L}_{h_c} : {\sf g}^* \mapsto {\sf g}$, defined by
$$ \langle \dot{\mu}, {\rm L}_{h_c} (\mu) \rangle = \frac{d}{dt} \vert_{t=0} 
h_c(\mu + t \dot{\mu} )~~.$$
Thus,
$$i_{X_H} \omega (v) = dH(v) = J^* dh_c(v) = d(h_c)_{J(m)} (T_m J v) =$$
$$\langle T_m J v, {\rm L}_{h_c} (J_m) \rangle = d \hat{J} ({\rm L}_{h_c} (J_m)) \cdot v = i_{({\rm L}_{h_c} \hat{J})_M} \omega (v)$$ 
such that $X_H (m)= ({\rm L}_{h_c} \circ J (m))_M(m)$. If $\gamma_t \in {\cal O}$ then ${\rm L}_{h_c} \gamma_t \subset {\sf g}$ and $a_t$ defined by $\dot{a}_t = T_e L_{a_t} ({\rm L}_{h_c} \gamma_t)$ is a curve in $G$. Considering $m_t = \Phi_{a_t} m_0$, then
$$ \dot{m}_t = \dot{\Phi}_{a_t} m_0 = (T_{a_t} L_{a_t^{-1}} \dot{a}_t)_M (m_t) =$$
$$({\rm L}_{h_c} \gamma_t)_M(m_t) = ({\rm L}_{h_c} \circ J(m_t))_M(m_t) = X_H (m_t)~~ \bigtriangledown. $$
In applications, to obtain the trajectory $m_t$ at $m_0$ are necessary several elements: \\
1. the orbit ${\cal O} = Ad^*_{a^{-1}} J (m_0) \subset {\sf g}^*$. \\
2. the trajectory $\gamma_t$ on ${\cal O}$ for the Hamiltonian system induced by $h_c$. \\
3. the curve $\xi_t \subset {\sf g}$, $\xi_t = {\rm L}_{h_c} (\gamma_t)$. \\
4. the trajectory $a_t \subset G$ provided by $\dot{a}_t = T_e L_{a_t} \xi_t$. \\
The result is then $m_t = \Phi_{a_t} (m_0)$. \\ \indent
It is important to remark that a relative equilibrium for a $G$-invariant Hamiltonian 
system moves on $G_\mu$-collective trajectories, corresponding to orbits $\gamma_t \subset 
{\sf g}^*$, degenerate in critical points of some $h_c$. Thus, in such situations the $G$-invariance implies $G_\mu$-collectivity. \\ \indent
In general, the action of $G$ on $(M, \omega)$ is not transitive, but the reduced phase-space $(\bar{M}, \bar{\omega})$ with respect to the action of $G$ is a homogeneous space  for a certain group $G_0$. This is the case for instance when: \\
- $M = T^* Q$ and $G$ acts transitively on $Q$. \\
- $M = T^* Q$ and $G$ acts freely on $Q$. \\
- $M=T^*G$. \\ 
{\it Theorem 6.} (Kirilov-Kostant-Souriau) Let $G$ be a Lie group, $L : G \times G \mapsto 
G$ the action of $G$ on $G$ by left translations, and $\Phi^\ell : G \times T^*G \mapsto 
T^*G$ the action induced by $L$ on $T^*G$ with the moment map $J^\ell : T^*G \mapsto 
{\sf g}^*$. Then (notation is explained in Appendix 1): \\
{\it i)} the reduced phase-space $(J^\ell)^{-1} (\mu) /G_\mu$ can be identified naturally with the orbit $G \cdot \mu = \{ Ad^*_{a^{-1}} \mu , a \in G \}$ of $\mu$ in ${\sf g}^*$. \\
{\it ii)} if $\tilde{R} : G \times G \mapsto G$ is the action to the right of $G$ on $G$, 
and $\tilde{\Phi}^r  : G \times T^*G \mapsto T^*G$ is the action induced by $\tilde{R}$ on 
$T^*G$, having the moment map $J^r : T^*G \mapsto {\sf g}^*$, then the $\Phi^\ell$-invariant 
Hamiltonians are $\tilde{\Phi}^r$-collective. \\
{\it Proof.}
{\it (i)}  For $\mu \in {\sf g}^*$,  $(J^\ell)^{-1} (\mu) = \{ (a, Ad_a^* \mu ), a \in G \}$. Let 
$p \in (J^\ell)^{-1} (\mu)$, $p = (a, Ad_a^* \mu )$. Because $\lambda_b p = (ba, Ad_a^* \mu)$, then $\lambda_b p 
\in (J^\ell)^{-1} (\mu)$ iff  $b \in G_\mu$. Thus, $G_\mu$ acts freely and properly on $(J^\ell)^{-1} (\mu)$, and
$$[(a, Ad_a^* \mu )] = \{ (ba, Ad_a^* \mu ), b \in G_\mu \} = Ad_a^* \mu \in G \cdot \mu~~. $$
{\it (ii)} Let $H : G \times {\sf g}^* \mapsto {\sf R}$ be the Hamilton function. Because 
$\lambda_b^* H(a, \mu) = H(ba, \mu)$, $H$ is $\lambda^*$-invariant only when it is 
independent of $a$, such that $H(a, \mu) = H(\mu) = 
h_c \circ J^r_{(a, \mu)}$, or $H = h_c \circ \tilde{J}^r$ $\bigtriangledown$. 
\\
{\it Example: the rigid body} \\
If $G=SO(3)$, then ${\sf g} \simeq {\sf g}^* \simeq {\sf R}^3$, and 
$\forall \mu \in {\sf g}^*$, $\mu = \sum_{i=1}^3 \mu_i f_i$, the orbit 
$G \cdot \mu$ is the sphere of radius $\vert \vec{\mu} \vert$, 
$\vec{\mu} = (\mu_1, \mu_2, \mu_3)$. The set $\{ f_i, i=1,2,3 \}$ of basis elements in 
${\sf g}^*$ is defined with respect to a set $\{ \xi_i, i=1,2,3 \}$, of basis elements in 
${\sf g}$, $[\xi_i, \xi_j]= \epsilon_{ijk} \xi_k$, such that 
$f_i ( \xi_k) = \delta_{ik}$. \\ \indent
Let us consider $H: T^*SO(3) \mapsto {\sf R}$ of the form $H=h_c \circ J^r$, with
$J^r (a, \mu) = - \mu$ and $h_c: {\sf g}^* \mapsto {\sf R}$,
\begin{equation}
h_c (\mu) = \frac{1}{2} \sum_{i=1}^3 \frac{ \mu_i^2}{I_i} ~~.
\end{equation} 
As $\hat{J}^r (\xi_i) : T^* G \mapsto {\sf R}$ has the expression
$ \hat{J}^r (\xi_i) = - \langle \mu , \xi_i \rangle = - \mu_i$,                
and the Poisson bracket satisfies $\{ \hat{J}^r (\xi), \hat{J}^r (\eta) \} = \hat{J}^r 
([\xi, \eta]) $ from equivariance, we get $\{ \mu_i, \mu_j \} = - \epsilon_{ijk} \mu_k$.  \\ \indent
The equations of motion are
\begin{equation}
\dot{\mu}_i = \{ \mu_i, h_c \} = \sum_{k=1}^3 \frac{ \mu_k}{I_k} \{ \mu_i, \mu_k \} =          
- \sum_{k=1}^3 \epsilon_{ikl} \frac{ \mu_k \mu_l}{I_k} ~~,
\end{equation}
or 
$$ \dot{\mu}_1 = \mu_2 \mu_3 (\frac{1}{I_3} - \frac{1}{I_2} )~,~  
\dot{\mu}_2 = \mu_1 \mu_3 (\frac{1}{I_1} - \frac{1}{I_3} )~,~
\dot{\mu}_3 = \mu_1 \mu_2 (\frac{1}{I_2} - \frac{1}{I_1} )~.
$$
Introducing $\omega_i = \langle f_i, {\rm L}_{h_c} (\mu) \rangle= \mu_i /I_i$ we also get
$$I_1 \dot{\omega}_1 = \omega_2 \omega_3 (I_2 -I_3)~,~ 
I_2 \dot{\omega}_2 = \omega_1 \omega_3 (I_3 -I_1)~,~
I_3 \dot{\omega}_3 = \omega_1 \omega_2 (I_1 -I_2)~,$$
Similar equations can be obtained considering the rigid body rotation as geodesic motion (Appendix 4). \\ \indent
When $I_1 > I_2 >I_3$ the critical points $\vec{\kappa} = (\kappa_1, \kappa_2, \kappa_3)$ of 
$h_c$ on $G \cdot \mu$, ($dh_c (T_\kappa G \cdot \mu ) =0$), are
$$ (0, \pm \vert \vec{\mu} \vert, o)~,~(\pm \vert \vec{\mu} \vert,0,0)~,
~(0,0,\pm \vert \vec{\mu} \vert)~.$$
For the trajectory in the chart $\Psi (T^* G)$ one obtains
$$ m_t = \rho_{a_t} (a_0, \mu_0 ) = (a_0 a_t^{-1} , Ad_{a_t^{-1}}^* \mu_0 ) $$
where $a_t$ is provided by $\dot{a}_t = T_e L_{a_t} \omega_t$, $\omega_t = L_{h_c} (\mu_t)$.   \\
As presented above, the Kirilov-Kostant-Souriau theorem concerns the case when on the 
configuration space $Q$ are defined two actions of the group $G$ which commute, $\Phi^r$ 
and $\Phi^\ell$, free and transitive, such that the action $\varphi : G \times Q \mapsto Q$, 
$\varphi_a = \Phi^r_a \Phi^\ell_a$ has a fixed point. In the example of the rigid body $\Phi^\ell$ corresponds to the rotation of the laboratory frame, 
while $\Phi^r$ to the rotation of the intrinsic frame. \\ 
{\it Theorem 7.} Let $\Phi : H \times Q \mapsto Q$ be a transitive action of the Lie group 
$H \subset Gl(V)$ on the manifold $Q$, and $\tau : H \times V \mapsto V$ a representation of $H$ in the linear space $V$. Assume that $ f: V \mapsto {\cal F}(Q)$ is a $H$-equivariant linear map, ($\Phi^*_{h^{-1}} f = f \circ \tau_h$, $\forall h \in H$), which defines a Hamiltonian action of the semidirect product $G = H \times V$ on $T^*Q$,
\begin{equation}
U^f_{(h,v)} \alpha_q = T^*_{\Phi_h q} \Phi_{h^{-1}} \alpha_q - df_v (\Phi_h q)~~,~~
(h,v) \in G~,~\alpha_q \in T^*Q~~, \label{act}
\end{equation}
with the equivariant moment map $J: T^*Q \mapsto {\sf h}^*+ {\sf V}^*$. If $\exists q_0 
\in Q$ such that $H_{q_0} = H_K$ (the action of $G$ on $T^*Q$ is transitive), then 
$J(T^*Q)$ is a single $G$-orbit in ${\sf g}^*$, specifically the $G$-orbit through the point 
$(0,K)$.  \\
{\it Proof.} The action (\ref{act}) has the form 
$$U^f_{(h,v)} \equiv t_{f_v} \hat{\Phi}_h ~~$$
where
$$\hat{\Phi}_h \alpha_q = T^*_{\Phi_h q} \Phi_{h^{-1}} \alpha_q~~,~~ t_{f_v} \alpha_q= \alpha_q - df_v (q) ~~.$$
If $j : T^*Q \mapsto {\sf h}^*$ denotes the moment map for the action of $H$ on $T^*Q$, 
$\pi : T^*Q \mapsto Q$ is the projection, and $v_{T^*Q} (\alpha_q) = - \pi^* d f_v (\alpha_q)$, 
then the moment map for the action $U^f$ is provided by
$$
\hat{J}_{\alpha_q}(\xi,v) = \hat{\j}_{\alpha_q} ( \xi) + \pi^* f_v (\alpha_q)~~. 
$$
Considering $f_v (q_0) \equiv \langle K, v \rangle$, $K \in V^*$, then $\forall q=\Phi_h(q_0)$ we get
$$ f_v(q)=  \langle K, \tau_{h^{-1}} v \rangle = \langle \tau^*_{h^{-1}} K, v \rangle $$
such that 
\begin{equation}
J_{\alpha_q} = (j_{\alpha_q} , \tau^*_{h^{-1}} K)~~, \label{mmap} 
\end{equation}
where $\alpha_q = \hat{\Phi}_h \alpha_{q_0}$, and $j_{\alpha_q} = j_{\hat{\Phi}_h 
\alpha_{q_0}} = (Ad^*_h)^{-1} j^0_p$, with $j^0_p = j_{\alpha_{q_0}}$,
$j^0 : T^*_{q_0}Q \mapsto {\sf h}^*$. In the chart on $T^*Q$ defined by
$$\Psi : T^* Q \mapsto Q \times T^*_{q_0} Q~~,~~\Psi (\alpha_q) =(q, \hat{\Phi}_{h^{-1}} 
\alpha_q ) \equiv (q,p)~~,   $$             
(\ref{mmap}) takes the form
\begin{equation}
J'_{(q,p)} = (Ad^*_{h^{-1}} j^0_p , \tau^*_{h^{-1}} K)~~.  
\end{equation}
Let $H_0 = \{ h \in H / \Phi_h q_0 = q_0 \}$ be the isotropy group of $q_0$, and 
${\sf h}_0$ its algebra. Thus, $\forall \xi \in {\sf h}_0$, $j^0_p (\xi)=0$, and in fact 
$j^0 : T^*_{q_0}Q \mapsto {\sf h}^*/ {\sf h}^*_0 \equiv \tilde{\sf h}^*_0$, 
$$ \tilde{\sf h}_0^* = \{ \mu \in {\sf h}^* / \mu (\xi) =0 ~\forall~ \xi \in {\sf h}_0 \}~~.$$
Let also 
$$H_K = \{ h \in H / \tau_{h^{-1}}^* K = K \}$$ 
be the isotropy group of $K$, and ${\sf h}_K$ its algebra. Presuming now that 
${\sf h}_0 = {\sf h}_K$, then ${\sf h}/ {\sf h}_0 = {\sf h}/{\sf h}_K$, 
$Q \simeq H \cdot K$, and as $H \subset Gl(V)$ the action of $G$ on $T^*Q$ is transitive. 
Thus, $J(T^*Q) $ is a covering space for the  $G$-orbit $G \cdot (0, K)$. \\ \indent
For $(\mu, K) \in {\sf h}^* + {\sf V}^*$ and $(h,v) \in H \times V$ we get
$$ Ad^*_{(h,v)^{-1}} ( \mu , K ) = (Ad^*_{h^{-1}} \mu + ( \tau^*_{h^{-1}} K ) \odot v,       
\tau^*_{h^{-1}} K)~~,$$
where $(h,v)^{-1} = (h^{-1}, - \tau_h^{-1}v)$, $Ad^*_{h^{-1}} ( K \odot v) = \tau^*_{h^{-1}}
K \odot \tau_h v$, and 
$(K \odot v) (\xi) \equiv \langle K , \xi_V (v) \rangle$ with
$$ \xi_V (v) = \frac{d}{dt} \vert_{t=0} \tau_{e^{t \xi}} (v) \in T_vV \simeq V ~~.$$
If $\pi_K : {\sf h} \mapsto {\sf h} / {\sf h}_K$, and $\mu_0 \in {\sf h}^*/{\sf h}_K^* 
\equiv \tilde{\sf h}^0_K$, then
$$  \pi^*_K (\mu - \mu_0) =0 \Leftrightarrow  \exists v \in V~,~\mu - \mu_0 = K \odot v ~~.$$
Thus, $ \forall j^0_p \in \tilde{\sf h}_K^*$, $\exists v \in V$ such that 
$K \odot v = j^0_p$,  $dim~ K \odot V = dim~ (H / H_0) = dim~ Q$, and
$$ J'_{(q,p)} = (Ad^*_{h^{-1}} j^0_p,  \tau^*_{h^{-1}} K) =  
 (Ad^*_{h^{-1}} (K \odot v) ,  \tau^*_{h^{-1}} K) = Ad^*_{(h^{-1},-v)} (0,K) ~~,$$
which shows that $J_{T^*Q}$ is the single orbit  $J_{T^*Q}=Ad^*_G (0, K)$ $\bigtriangledown$. \\ \indent
One should note that when $Q$ is the orbit $Q=H \cdot K \subset {\sf V}^*$, then $T^*Q$ is mapped by $J$ on the $G$-orbit  $G \cdot (0, K)$, and any Hamiltonian on $M = T^*Q$ is $G$-collective. Moreover, if $(M, \omega)$ is an arbitrary phase-space, and $J : M \mapsto V^*$, then the action $\Phi$ defines a physical collective model. The most general coadjoint orbit of the semidirect product $G = H \times V$ appears as reduced phase-space with respect to the action to the right of $H_K$ on $T^*H$.
\subsection{Examples}
1. {\it The rigid body in external field} \\
Let $Q= SO(3, {\sf R}) =H$ and $V={\sf R}^3$. For the action to the right of $G = H \times V$ on $T^*Q$ the moment map is 
$$ J^r (h, \mu) = ( - \mu, \tau^*_{h^{-1}} K )~~.$$
If $K \in {\sf V}^* \simeq {\sf R}^3$ is a constant force field, and 
$x \in V \simeq {\sf R}^3$ is the center of mass position vector, then the Hamiltonian on $T^*SO(3)$ is $H = h_c \circ J^r$, with
$$ h_c ( \mu, K) = \frac{1}{2} \sum_{i=1}^3 \frac{\mu_i^2}{2I_i} - \langle K, x \rangle~~.$$
2. {\it The liquid drop model} \\
Let $V = \{ w \in Gl(3, {\sf R}) / w^T=w, {\rm det} ~ w >0 \} \simeq {\sf R}^6$ be the space of the quadrupole moments  $w_{ij} = \sum_{n=1}^N x_{ni} x_{nj}$, $i,j=1,2,3$ for a system of $N$ particles. An action $\tau$ of $H = Sl(3)$ on $V$, $\tau : Sl(3) \times V \mapsto V$ 
$$\tau_h w = h w h^T~~,h \in Sl(3) $$
is induced by its action  ${\bf x}_n \mapsto h {\bf x}_n$ on ${\sf R}^{3N}$, and 
$Q = H \cdot K$, $K \in {\sf V}^*$, is the 5-dimensional configuration space for the liquid 
drop.  \\ \indent
Let $G$ be the 14-dimensional group $G = H \times V \equiv CM(3)$,
$$CM(3) = \{ (h, w) = \left[ \begin{array}{cc} h & 0  \\ 
wh &  (h^T)^{-1}   \end{array} \right] ~,~~h \in SL(3), w \in V \simeq {\sf R}^6 \} ~~.
$$    
Thus, 
$${\sf cm}(3) = \{ (\xi, \eta) = \left[ \begin{array}{cc} \xi & 0  \\ 
\eta &  - \xi^T   \end{array} \right] ~,~~\xi \in {\sf sl}(3), \eta \in {\sf V} \} \subset {\sf sp}(3, {\sf R}) \subset {\sf gl}(3, {\sf R})~~,
$$    
and
$$
{\sf cm}(3)^* = \{ (\mu, \nu)^\flat~,~tr \mu =0~,~\nu=\nu^T / \mu \in {\sf sl}(3), \nu \in {\sf V}^* \} 
$$ 
with
$$
\langle ( \mu, \nu)^\flat , \left[ \begin{array}{cc} \xi & 0  \\ 
\eta &  - \xi^T   \end{array} \right]  \rangle \equiv \frac{1}{2} [ tr (\mu^T \xi) + tr (\nu \eta)]~~.
$$
For the element $(\alpha L_3, \beta I)^\flat$, $L_3 = e_{12}-e_{21}$, (Appendix 2), one obtains 
$$Ad^*_{(h,w)^{-1}} ( \alpha L_3, \beta I)^\flat = (\alpha (h^T)^{-1} L_3 h^T - 2 \beta w h h^T + \frac{2}{3} \beta Tr(whh^T) , \beta h h^T )^\flat
$$ 
such that when $\alpha \ne 0$ the stability group $G_{(\alpha, \beta)}$ of $(\alpha L_3, \beta I)^\flat$ is
$$G_{(\alpha, \beta)} = \{ (SO(2), w_0 I), w_0 \in {\sf R} \}~~, $$
while for $\alpha =0$ 
$$G_{(0, \beta)} = \{ (SO(3), w_0 I), w_0 \in {\sf R} \}~~. $$
Thus, as $CM(3)$ is 14-dimensional, when $\alpha \ne 0$ the coadjoint orbits are 12-dimensional,  and 10-dimensional when $\alpha =0$. The moment map $J : T^*Q \mapsto {\sf g}^*$ is a diffeomorphism from the phase space $M=T^*Q$ to the coadjoint orbit of the 
element $(0, \beta I)^\flat$.  \\   
 {\it 3. The relativistic particle} \\
If $H = SO(3,1)$ is the Lorentz group, and $V = {\sf R}^{3,1}$ is the Minkowski space, then
$G = H \times V$ is the Poincar\'e group. The Lie algebra ${\sf so}(3,1)$ is generated by six 
$4 \times 4$ real matrices, $\{ J_i, K_i, i=1,2,3 \}$, with nonvanishing elements 
$(J_i)_{kl} = \epsilon_{ikl}$, $i,k,l=1,2,3$, $(K_i)_{\mu \nu} = \delta_{\mu 0} \delta_{i \nu} + \delta_{\nu 0} \delta_{ i \mu}$, 
$\mu, \nu = 0,1,2,3$, and commutation relations
$[J_i, J_k] = - \epsilon_{ikm} J_m$, $[K_l, K_m] = \epsilon_{lmi} J_i$, $ [J_m, K_i] = - \epsilon_{mil} K_l$. \\ \indent
As\footnote{${\sf sl}(2,{\sf C}) \simeq {\sf so}(3, {\sf C}) \simeq
{\sf sp}(2, {\sf C})$, with the real forms
${\sf sl}(2,{\sf R}) \simeq {\sf so}(2,1) \simeq {\sf su}(1,1) \simeq
{\sf sp}(2, {\sf R})$ (split)
and ${\sf su}(2) \simeq {\sf so}(3, {\sf R})$ (compact).} ${\sf so} (3,1)
\simeq {\sf sl}(2, {\sf C})$, an element $\xi = {\bf \alpha} \cdot {\bf K} + {\bf
\beta} \cdot {\bf J} \in {\sf so}(3,1)$, ${\bf \alpha}, {\bf \beta} \in
{\sf R}^3$, corresponds to $x=( {\bf \alpha} +i {\bf \beta}) \cdot {\bf
\sigma}/2 \in {\sf sl} (2, {\bf C})$, by ${\bf \sigma} \equiv (\sigma_1,
\sigma_2, \sigma_3) $ denoting the Pauli matrices. \\ \indent
An element $F \in {\sf g}^*$ can be expressed in the form         
$$ F = p_0 X_0^* - {\bf p} \cdot {\bf X}^* + {\bf \kappa} \cdot {\bf K}^* + {\bf s} \cdot {\bf J}^*~~,$$
where the basis $(X_0^*, {\bf X}^*, {\bf K}^*, {\bf J}^*) $ is defined by
$X_\mu^* (X_\nu) = \delta_{\mu, \nu}$ , $K_i^*(K_l) = J_i^*(J_l) = \delta_{il} $. Denoting $f \equiv (p_0, {\bf p}, {\bf \kappa}, 
{\bf s})$, the orbit $G \cdot F$ contains $f_0 \equiv (m_0c, {\bf 0},
{\bf 0}, {\bf s}_0)$, where $m_0$ and ${\bf s}_0$ are the
"intrinsic" mass and angular momentum. As $G_{f_0} \simeq {\sf R} \times SO(2, {\sf R})$, (time translations $\times$ rotations around ${\bf s}_0$),
if ${\bf s}_0 \ne 0$, and $G_{f_0} \simeq {\sf R} \times SO(3, {\sf R})$ if ${\bf s}_0=0$, for massive particles the orbits $G \cdot F \simeq 
G / G_{f_0}$ are 8, respectively 6 dimensional.   
\section{Appendix 1: Adjoint group actions } 
Let $G$ be a Lie group, and $M = T^*G$. The left and right translations on $G$ are defined 
by $L_a (h) =  a h$, and $R_a (h) =h a$, respectively, while $\tilde{R}_a = R_{a^{-1}}$ is the 
action to the right. \\
For $\xi_h \in T_hG$ one gets
$$ T_h L_a \xi_h \in T_{ah} G~~,~~  T_h R_a \xi_h \in T_{ha} G~~,$$
while for $\alpha \in T^*_h G$, 
$$ T_{ah} L_{a^{-1}}^* \alpha_h \in T_{ah}^* G~~,~~  T_{ha^{-1}} R_a^* \alpha_h \in T_{ha^{-1}}^* G~~.$$
One denotes by $\Phi^\ell_a = T L_{a^{-1}}^*$ and $\Phi^r_a = TR^*_a$ 
the actions induced by $L$ and $\tilde{R}$ on $M$. 
 Let $\Psi : M \mapsto G \times {\sf g}^*$  be the map 
$$ \Psi(\alpha_a) = (a, T_e L^*_a \alpha_a )= (a, \Phi^\ell_{a^{-1}} \alpha_a) \in G \times {\sf g}^*~~.$$
In the chart $\Psi(M)$, (of intrinsic coordinates), the actions $\Phi^\ell$ and $\Phi^r$ 
are represented by the actions denoted $\lambda$ and $\rho$, respectively,  
$$ \Psi( \Phi^\ell_a \alpha_h ) \equiv \lambda_a \Psi( \alpha_h) ~~,~~
\Psi( \Phi^r_a \alpha_h ) \equiv \rho_a \Psi( \alpha_h) ~~,$$
provided by
$$ \lambda_a (h, \mu) = (ah, \mu)~~,~~\rho_a (h, \mu) = (h a^{-1}, Ad^*_{a^{-1}} \mu) ~~,$$
where 
$$Ad^*_{a^{-1}} = \Phi^r_a \Phi^\ell_{a^{-1}} = T_e (L_{a^{-1}} R_a)^*~~.$$
The projection on $TG$ of the infinitesimal generators for these actions are
$$ \xi^\ell_G (a) = T_e R_a \xi~~,~~ \xi^r_G(a)= - T_e L_a \xi~~.$$
such that the moment maps at $\alpha_a = \Phi^\ell_a \mu$ are
$$ J^\ell_{\alpha_a} \cdot \xi = \alpha_a (\xi^\ell_G(a)) = T_e R^*_a \alpha_a (\xi) = (Ad^*_{a^{-1}} \alpha) \cdot \xi~~,$$
$$ J^r_{\alpha_a} \cdot \xi = \Phi^\ell_a \mu (\xi^r_G(a)) = - (\Phi^\ell_{a^{-1}} \Phi^\ell_a \mu) (\xi) = - \mu \cdot \xi~~,$$
or
$$ J^\ell (a, \mu) = Ad^*_{a^{-1}} \mu ~~,~~J^r (a, \mu) = - \mu~~.$$
These maps are equivariant, as
$$ J^\ell (\lambda_a(h, \mu))= Ad^*_{a^{-1}} J^\ell (h, \mu)~~,~~
J^r (\rho_a(h, \mu))= Ad^*_{a^{-1}} J^r (h, \mu)~~.$$ 
\section{Appendix 2: The classical semisimple Lie algebras} 
The root space decomposition ${\sf g }= {\sf h} + \sum_{\alpha \ne 0} {\sf g}_\alpha$ of a complex semisimple Lie 
algebra ${\sf g}$ with respect to the adjoint representation ($ad_x y = [x,y]$) of a Cartan subalgebra ${\sf h}$ 
provides the Weyl basis $\{ h_i, e_\alpha, i=1, {\rm rank}({\sf g}), \alpha \in {\sf h}^* \}$, with 
$$ [h, e_\alpha ] = \alpha (h) e_\alpha ~~,~~ [ e_\alpha , e_{ - \alpha} ] = ( e_\alpha , e_{- \alpha} ) 
h_\alpha~~.$$  
Here $(x,y) \equiv Tr(ad_x ad_y)$ is the (nondegenerate) Cartan-Killing form, $\alpha (h) \equiv (h_\alpha, h)$, 
$(\alpha, \beta) \equiv ( h'_\alpha, h'_\beta)$. If $( \alpha_1, \alpha_2,..., \alpha_l)$ is an ordered basis in 
${\sf h}^*$, then $\beta = \sum_i c_i \alpha_i $ is positive ($\in \Sigma$) if the first $c_i \ne 0$ is positive, 
and simple if it is positive, but not a sum of positive roots. For the simple roots $\alpha_i$, $\alpha_j$, the 
elements of the Cartan matrix $A_{ij} = 2 (\alpha_i, \alpha_j) / (\alpha_i, \alpha_i)$ are integers, so that the 
real span of the Weyl basis is the normal (or split) real form of ${\sf g}$.
If $V$ is a finite dimensional vector space, ${\sf gl}(V)$ is reductive, and
$[{\sf gl}(V), {\sf gl}(V)]= {\sf sl}(V)$ is semisimple. When $V={\sf C}^n$,
a basis of ${\sf gl}({\sf C}^n) \equiv {\sf gl}(n, {\sf C})$ is provided by
the set of $n^2$ real matrices $\{ e_{pq}, p,q=1,n \}$,
with the single non-vanishing element, equal to 1, in the row $p$ and column $q$, $(e_{pq})_{ij} = \delta_{pi} \delta_{qj}$, 
and $[e_{pq}. e_{rs} ] = \delta_{qr} e_{ps} - \delta_{ps} 
e_{rq}$, $( e_{pp}, e_{qq}) = \sum_{ij} (\delta_{ip} - \delta_{jp}) (\delta_{iq} - \delta_{jq}) = 2n \delta_{pq} 
-2$. This basis can be represented using a set of $n$ boson operators $\{ b_p^\dagger, b_p, p=1,n \}$, 
$[b_p, b_q^\dagger] = \delta_{pq}$, as $e_{pq} = b_p^\dagger b_q$, or a set of $n$ fermion operators,
$\{ c_p^\dagger, c_p, p=1,n \}$, $ \{ c_p, c_q^\dagger \} = \delta_{pq}$, as $e_{pq} = c_p^\dagger c_q$. 
\subsection{ Semisimple Lie algebras of type $A_n:{\sf sl}(n+1,{\sf C})$}
${\sf sl}(n+1, {\sf C}) = \{ \xi \in {\sf gl}(n+1,{\sf C})/ Tr \xi =0 \}$. If $e_{pq} \in {\sf gl} (n+1, {\sf C} )$, $(e_{pq})_{ik} = \delta_{ip} 
\delta_{kq}$, the the diagonal matrices $h_m$, $m=1,n$, with 
$$h_m = \sum_{i=1}^{n+1} c^i_m e_{ii}~~,~~ \sum_{i=1}^{n+1} c^i_m=0~~,~~ \sum_{i=1}^{n+1} c^i_m c^i_{m'} = \delta_{mm'}$$
satisfy $(h_m,h_{m'}) =2(n+1) \delta_{mm'}$, and provide an orthogonal basis in the Cartan subalgebra ${\sf h}$ of ${\sf sl}(n+1,{\sf C})$. 
For these elements $[h_m, e_{pq}]= \alpha_{pq}(m) e_{pq}$ with \\
$$\alpha_{pq}(m)=- \alpha_{qp}(m)= \sum_{i=1}^{n+1} c^i_m (\delta_{ip} - \delta_{iq}) =c^p_m -c^q_m ~~.$$
Denoting $\beta_p \equiv \alpha_{pp+1}$, we get $\alpha_{pq}=\sum_{i=p}^{q-1} \beta_i$. These results provide: \\
- the set of roots: \\
$$ \Delta = \{ \alpha_{pq}, p \ne q \}$$
with $(n+1)^2-(n+1)$ elements. \\
- the set of positive roots: \\ 
$$ \Sigma = \{ \alpha_{pq}, p < q \}$$
with $n(n+1)/2$ elements. \\
- the set of simple roots: \\ 
$$ \Pi = \{ \beta_p=\alpha_{pp+1}, p=1,n \}$$
with $n$ elements. A particular basis in ${\sf h}$ consists of the elements $g_m = e_{mm}-e_{m+1m+1}$, $m=1,n$. For this basis we get 
$$[g_m, e_{pq}]= (g_m, h'_{\alpha_{pq}}) e_{pq}~~,~~ 
[e_{pq}, e_{qp}] =e_{pp}-e_{qq} = \sum_{i=p}^{q-1}g_i = 2(n+1) h'_{\alpha_{pq}}$$
where
$$h'_{\alpha_{pq}}= \frac{1}{2(n+1)} \sum_{i=p}^{q-1} g_i~~,~~(g_m, h'_{\alpha_{pq}}) = \delta_{mp}-\delta_{mq}+ \delta_{m+1q}-\delta_{m+1p}$$
As $( \beta_p, \beta_q) = (h'_{\beta_p}, h'_{\beta_q})= (g_p, g_q) / 4 (n+1)^2 = ( 2 \delta_{pq} - \delta_{p q+1} - \delta_{p+1q})/ 2(n+1)$, we get the Cartan matrix
$$A_{mm'} = 2 \frac{(\beta_m,\beta_{m'})}{(\beta_m,\beta_m)} =2 \delta_{mm'}- \delta_{mm'+1}-\delta_{m+1m'}~~,$$
and the Dynkin diagram\footnote{The circles $\{ \circ^i \}$ correspond to the simple roots $\{ \beta_i \}$ and the number of lines 
between $\circ^i$ and $\circ^j$ is equal to $A_{ij}A_{ji} = 4 \cos^2 \theta_{ij}$, where $\theta_{ij}$ is the angle between   $\beta_i$ and $\beta_j$. 
Whenever $(\beta_i, \beta_i) < (\beta_j, \beta_j)$, the lines get an arrow pointing towards $\circ^i$
\cite{wolf}.}  \cite{jacob} presented
in Figure 1.  

\begin{figure}
\begin{picture}(0,0)(-150,60)
\put(-97,110){1}
\put(-100,100){$\bigcirc$}
\put(-88,103){\line(1,0){27}}
\put(-57,110){2}
\put(-60,100){$\bigcirc$}
\put(-48,103){\line(1,0){27}}
\put(-19,110){3}
\put(-22,100){$\bigcirc$}
\put(-10,103){$\dots \dots$}
\put(15,110){n-1}
\put(16,100){$\bigcirc$}
\put(28,103){\line(1,0){27}}
\put(58,110){n}
\put(56,100){$\bigcirc$}
\end{picture}
{\small Figure 1. Dynkin diagram for the $A_n$ algebras }
\end{figure}
The compact form ${\sf su}(3)$ of ${\sf sl}(3, {\sf C})$ is presented in detail in
\cite{su3}.

\subsection{ Semisimple Lie algebras of type $B_n:{\sf so}(2n+1,{\sf C})$}
${\sf so}(2n+1, {\sf C}) = \{ \xi \in {\sf gl}(2n+1,{\sf C})/ \xi^T = - \xi \}$, with basis provided 
by $n(2n+1)$ independent matrices $\xi_{pq} = e_{pq} - e _{qp}$, $p,q= -n,...,-1,0,1,...n$. The
real span of this basis generates the compact real form ${\sf so}(2n+1, {\sf R})$, but to obtain 
the Weyl basis the representation should be changed to $w \xi w^{-1}$, with 
$w_{ij} = [ (1+i) \delta_{ij} + (1-i) \delta_{i-j}]/2$, ($w^{-1} = w^\dagger = w^*$). In the new
representation   ${\sf so}(2n+1, {\sf C})$ is generated by the elements
$$ \{ f_{pq} = -f_{-q-p}=e_{pq}-e_{-q-p}/ p,q=-n,...,-1,0,1,...,n \}~~,$$
while the real span of $f_{pq}$ generates ${\sf so} (n+1,n)$.  \\ \indent
As $[f_{pq}, f_{kl} ] = \delta_{qk} f_{pl} - \delta_{lp} f_{kq}
+\delta_{p-k} f_{-lq} + \delta_{-ql} f_{k-p}$ we get
$$[f_{pp}, f_{kl}] = \alpha_{kl}(p) f_{kl}~~, $$
with $\alpha_{kl}(p) = \delta_{pk} - \delta_{pl} +\delta_{p-l} 
-\delta_{p-k}$, and
$$[f_{kl},f_{lk}]= f_{kk} - (1- 2 \delta_{k-l}) f_{ll}~~. $$
These commutation relations provide: \\
- the set of roots: \\
$$ \Delta = \{ \alpha_{kl}(p), k >- l, k \ne l \}$$
with $2n^2$ elements. \\
- the set of positive roots: \\ 
$$ \Sigma = \{ \alpha_{kl}(p)=\delta_{pk}-\delta_{pl}, l >k >0 \} \cup $$
$$
\{ \alpha_{kl}(p)=- \delta_{p-k}-\delta_{pl}, -l <k<0  \} \cup
\{ \alpha_{0l}(p)=-\delta_{pl}, 1 \le l \le n \}
$$
with $n^2$ elements. \\
- the set of simple roots: \\ 
$$ \Pi = \{ \beta_0 = \alpha_{01} \} \cup \{ \beta_k=\alpha_{kk+1}, k=1,n-1 \} $$
with $n$ elements. In terms of the simple roots
$$\alpha_{kl}= \sum_{i=\vert k \vert}^{l-1} \beta_i + 2 \sum_{i=0}^{\vert k \vert -1} \beta_i~~{\rm if}~~ 
-l<k<0 ~~,$$
$\alpha_{0l}= \sum_{i=0}^{l-1} \beta_i $, and
$$\alpha_{kl}= \sum_{i=k}^{l-1} \beta_i ~~{\rm if}~~l>k>0~~.$$
Because
$$h'_{\alpha_{0l}} = - \frac{f_{ll}}{2(2n-1)}~~,~~ 
h'_{\alpha_{kk+1}} =  \frac{f_{kk}-f_{k+1~k+1}}{2(2n-1)}~~,~~.$$
and  
$$( f_{pp}, f_{qq}) = \sum_{-l<k =-n}^n  \alpha_{kl}(p) \alpha_{kl}(q) =
2 (2n-1) \delta_{pq} $$
we get
$$(\beta_0, \beta_0) = \frac{1}{2(2n-1)}~~,~~
(\beta_k, \beta_k) = -2(\beta_{k-1}, \beta_k)= \frac{1}{2n-1}  ~,k=1,n-1~~,~~$$
such that the angles between the simple roots are specified by
$\cos \varphi_{01} = -1 / \sqrt{2}$ and $\cos \varphi_{kk+1} = -1 /2$, $k>0$. The
Dynkin diagram associated to the Cartan matrix $[2 (\beta_i, \beta_j) /
(\beta_i, \beta_i ) ]$ is represented in Figure 2.  \\ \indent
\begin{figure}
\begin{picture}(0,0)(-150,60)
\put(-85,110){0}
\put(-88,100.5){$\bigcirc$}
\put(-78,100){$\Longleftarrow$}
\put(-57,110){1}
\put(-60,100){$\bigcirc$}
\put(-48,103){\line(1,0){27}}
\put(-19,110){2}
\put(-22,100){$\bigcirc$}
\put(-10,103){$\dots \dots$}
\put(15,110){n-2}
\put(16,100){$\bigcirc$}
\put(28,103){\line(1,0){27}}
\put(54,110){n-1}
\put(56,100){$\bigcirc$}
\end{picture}
{\small Figure 2. Dynkin diagram for the $B_n$ algebras }
\end{figure}
In the case of ${\sf so}(5. {\sf C})$, if $\Sigma = \{ \alpha_{-12}, \alpha_{01},
\alpha_{02}, \alpha_{12} \}$, then $\Pi = \{ \alpha_{01}, \alpha_{12} \}$.
Some applications with ${\sf so}(5)$ as a subalgebra of ${\sf so}(8)$  are
presented in \cite{so51, so52}.
\subsection{ Semisimple Lie algebras of type $C_n:{\sf sp}(2n,{\sf C})$}
${\sf sp}(2n, {\sf C}) = \{ \xi \in {\sf gl}(2n,{\sf C})/ \xi^TJ = -J \xi  \}$, where $J$ has the $n \times n$ block form 
$$J= \left[ \begin{array}{cc} 0 & I  \\ 
-I &  0   \end{array} \right] ~~.
$$
Because any element $\xi \in {\sf sp}(2n, {\sf C})$ can  be expressed as
$$
\xi =  \left[ \begin{array}{cc} a & b  \\ 
c &  - a^T   \end{array} \right]~~,~~ b=b^T~,~c=c^T~~,
$$
a basis in ${\sf sp}(2n, {\sf C})$ is provided by $n^2$ independent matrices $A$,
$$
A_{ij} =  \left[ \begin{array}{cc} e_{ij} & 0  \\ 
0 &  - e_{ji}   \end{array} \right]~~,
$$
$n(n+1)/2$ independent matrices $B$, 
$$
B_{ij} =  \left[ \begin{array}{cc} 0 & e_{ij}  \\ 
0 &  0   \end{array} \right] + \left[ \begin{array}{cc} 0 & e_{ji}  \\ 
0 &  0   \end{array} \right]~~,$$
and 
$n(n+1)/2$ independent matrices $C$,
$$
C_{ij} =  \left[ \begin{array}{cc} 0 & 0  \\ 
e_{ij} &  0   \end{array} \right] + \left[ \begin{array}{cc} 0 & 0  \\ 
e_{ji} &  0   \end{array} \right]~~.$$
The commutation relations are specified by
$$[A_{ii}, A_{pq}]= (\delta_{ip}-\delta_{iq}) A_{pq}~~,$$  
$$[A_{ii}, B_{pq}]= (\delta_{ip}+\delta_{iq}) B_{pq}~~,$$
$$[A_{ii}, C_{pq}]=- (\delta_{ip}+\delta_{iq}) C_{pq}~~,$$
$$[B_{pq},C_{pq}] = (2 \delta_{pq}+1) A_{pp} +A_{qq}~~,$$
$$[A_{pq},A_{qp}]=A_{pp}-A_{qq}~~,$$ 
with $i,p,q=1,n$. Denoting $\alpha_{pq}(i)= \delta_{ip}-\delta_{iq}$, $p>q$ and  $\alpha'_{pq}(i)= (\delta_{ip}+\delta_{iq})$, $p \ge q$, we get the set of roots: \\
$$ \Delta = \{ \alpha_{pq}, - \alpha_{pq},~ p,q=1,n / p >q \} \cup 
\{ \alpha'_{pq}, - \alpha'_{pq},~ p,q=1,n / p \ge q \}~~, $$
with $2n^2$ elements. \\
- the set of positive roots: \\ 
$$ \Sigma = \{ \alpha_{pq},~ p,q=1,n / p <q \} \cup 
\{ \alpha'_{pq}, ,~ p,q=1,n / p \le q \}~~, $$
with $n^2$ elements. \\
- the set of simple roots: \\ 
$$ \Pi = \{\beta_p= \alpha_{pp+1}, p=1,n-1 \} \cup \{\beta_n= \alpha'_{nn} \} $$
with $n$ elements. In terms of the simple roots
$$\alpha_{pq}= \sum_{i=p}^{q-1} \beta_i~~,~~\alpha'_{qq}=\beta_n+ 2 \sum_{i=q}^{n-1} \beta_i~~,~~\alpha'_{pq}=(1-\delta_{pq}) \alpha_{pq}+ \alpha'_{qq}~~. $$
The Cartan subalgebra is generated by $n$ independent diagonal matrices,
$$h'_{\beta_p} =  \frac{A_{pp}-A_{p+1p+1}}{4(n+1)}~,~p<n~~,~~ 
h'_{\beta_n} = \frac{A_{nn}}{2(n+1)}~~.$$
Because 
$$( A_{ii}, A_{jj}) = 2 \sum_{p>q}  \alpha_{pq}(i) \alpha_{pq}(j) 
+ 2 \sum_{p \geq q} \alpha'_{pq}(i) \alpha'_{pq}(j) =
4 (n+1) \delta_{ij} $$
we get
$$(\beta_p, \beta_q) = \frac{2 \delta_{pq} - \delta_{pq+1} - \delta_{p+1q}}{4(n+1)} 
~~,~~
(\beta_p, \beta_n) = - \frac{\delta_{pn-1}}{2(n+1)}~~,p<n$$
and $(\beta_n, \beta_n) = 1/(n+1)$. The Dynkin diagram associated with the matrix 
$[2 (\beta_i, \beta_j) / (\beta_i, \beta_i )] $ is represented in Figure 3.   
\begin{figure}
\begin{picture}(0,0)(-150,60)
\put(-97,110){1}
\put(-100,100){$\bigcirc$}
\put(-88,103){\line(1,0){27}}
\put(-57,110){2}
\put(-60,100){$\bigcirc$}
\put(-48,103){\line(1,0){27}}
\put(-19,110){3}
\put(-22,100){$\bigcirc$}
\put(-10,103){$\dots \dots$}
\put(17,110){n-1}
\put(18,100){$\bigcirc$}
\put(28,100){$\Longleftarrow$}
\put(48,110){n}
\put(46,100){$\bigcirc$}

\end{picture}
{\small Figure 3. Dynkin diagram for the $C_n$ algebras }
\end{figure}

\subsection{ Semisimple Lie algebras of type $D_n:{\sf so}(2n,{\sf C})$} 
${\sf so}(2n, {\sf C}) = \{ \xi \in {\sf gl}(2n,{\sf C})/ \xi^T = - \xi \}$ 
Considering the notation introduced above for $B_n$, in the Weyl basis  
${\sf so}(2n, {\sf C})$ is generated by the elements 
$$\{ f_{pq} = -f_{-q-p}=e_{pq}-e_{-q-p}/ p,q=-n,...,-1,1,...,n \}~~,$$
having as real span ${\sf so}(n,n)$. Because ${\sf h} = \{ f_{pp}. p=1,n \}$, and
$$[f_{pq}, f_{kl} ] = \delta_{qk} f_{pl} - \delta_{lp} f_{kq}
+\delta_{p-k} f_{-lq} + \delta_{-ql} f_{k-p}~~,$$ we get
$[f_{pp}, f_{kl}] = \alpha_{kl}(p) f_{kl} $
with $\alpha_{kl}(p) = \delta_{pk} - \delta_{pl} +\delta_{p-l} -\delta_{p-k}$. 
Also 
$$[f_{kl},f_{lk}]= f_{kk} - (1- 2 \delta_{k-l}) f_{ll}~~. $$
These relations provide: \\ 
- the set of roots: \\
$$ \Delta = \{ \alpha_{kl}(p), k >- l, k \ne l/ k,l=-n,...,-1,1,...n \}$$
with $2n(n-1)$ elements. \\
- the set of positive roots $\alpha_{kl}$ 
with $k > \vert l \vert $ or $l > \vert k \vert$, \\ 
$$ \Sigma = \{ \alpha_{kl}(p)=\delta_{pk}-\delta_{pl}, l >k >0 \} \cup $$
$$
\{ \alpha_{kl}(p)=- \delta_{p \vert k \vert}-\delta_{pl}, -l<k<0 \} 
$$
with $n(n-1)$ elements. \\
- the set of simple roots: \\ 
$$ \Pi = \{ \beta_0 = \alpha_{-12} \} \cup \{ \beta_k=\alpha_{kk+1}, k=1,n-1 \} $$
with $n$ elements. In terms of the simple roots
$$\alpha_{kl}= \sum_{i=k }^{l-1} \beta_i~~,~~ l > k >0~~,$$
$$\alpha_{- \vert k \vert \vert k \vert+1 }= \beta_0+ \beta_1+\beta_k +2 
\sum_{i=2}^{\vert k \vert-1} \beta_i ~~,~~k=1,n-1~~,$$
and $\alpha_{-\vert k \vert l} = \alpha_{- \vert k \vert \vert k \vert+1} + \alpha _{\vert k \vert +1 l} $, 
$l > \vert k \vert$.  
Because 
$$ h'_{\alpha_{-12}} = \frac{ -f_{11} -f_{22}}{4(n-1)}~~,~~
h'_{\alpha_{kl}} = \frac{ f_{kk} -f_{ll}}{4(n-1)}~, l>k>0~~,
$$
and
$$( f_{pp}, f_{qq}) = \sum_{-l<k=-n}^n  \alpha_{kl}(p) \alpha_{kl}(q) =
4 (n-1) \delta_{pq} $$
we get $(\beta_0, \beta_1) =0$, 
$$(\beta_0, \beta_0) = (\beta_k, \beta_k) =
-2 (\beta_0, \beta_2) =-2 (\beta_k, \beta_{k+1})= \frac{1}{2(n-1)}~~, k=1,n-1~~$$
such that the angles between the simple roots are specified by $\cos \varphi_{01} = 0 $,
$\cos \varphi_{02} = -1/2 $ and $\cos \varphi_{kk+1} = -1 /2$ for $k \ge 1$. The Dynkin diagram associated with the matrix $[2 (\beta_i, \beta_j) / (\beta_i, \beta_i )] $ is represented in Figure 4.   
\begin{figure}
\begin{picture}(0,0)(-150,50)
\put(-97,142){1}
\put(-97,82){0}
\put(-100,130){$\bigcirc$}
\put(-60,103){\line(-1,-1){27}}
\put(-100,70){$\bigcirc$}
\put(-60,103){\line(-1,1){27}}
\put(-57,110){2}
\put(-60,100){$\bigcirc$}
\put(-48,103){\line(1,0){27}}
\put(-19,110){3}
\put(-22,100){$\bigcirc$}
\put(-10,103){$\dots \dots$}
\put(15,110){n-2}
\put(16,100){$\bigcirc$}
\put(28,103){\line(1,0){27}}
\put(53,110){n-1}
\put(56,100){$\bigcirc$}

\end{picture}
{\small Figure 4. Dynkin diagram for the $D_n$ algebras }
\end{figure}

Let $c^\dagger_i, c_i$ be the fermion creation and annihilation operators for a quantum many-body system described in terms of $n>1$ 
single-particle states, specified by $i=1,n$. Then 
$$[c_i^\dagger c_i, c^\dagger_j c^\dagger_k ] = (\delta_{ij}+ \delta_{ik}) c^\dagger_j c^\dagger_k~~,~~ [c_i^\dagger c_i, c_j c_k ] = -(\delta_{ij}+ \delta_{ik}) c_j c_k~~,$$
and $[c_i^\dagger c_i, c^\dagger_j c_k ] = (\delta_{ij}- \delta_{ik}) c^\dagger_j c_k$, so that the set 
$S=\{ c^\dagger_i c^\dagger_j, c_i c_j, c^\dagger_i c_j/ i,j =1,n \}$  generates an ${\sf so}(2n)$ algebra. For a two-state system 
${\sf so} (4, {\sf C}) \simeq {\sf sl} (2, {\sf C})+{\sf sl} (2, {\sf C}) $ is generated by 
$$J_+ = c^\dagger_1 c_2~~,~~J_0 = \frac{1}{2} (c^\dagger_1 c_1 - c^\dagger_2 c_2),~~J_- = c^\dagger_2 c_1 $$  
and 
$$P_+ = c^\dagger_1 c^\dagger_2~~,~~P_0 = \frac{1}{2} (c^\dagger_1 c_1 + c^\dagger_2 c_2-1),~~P_- = c_2 c_1~~. $$  
An application of the coherent states generated by $S$ can be found in
\cite{av}.
\section{Appendix 3: Symplectic actions and invariance} 
Let $(M, \omega )$ be a symplectic manifold, $\Phi_a$ a symplectic action of the Lie group $G$ on $M$, and $X_H$ a Hamiltonian 
field on $M$, $i_{X_H} \omega = dH$. The equations of motion are called invariant to the action of $G$ if
$$ \Phi_a^* X_H = X_H ~~,~~ \forall a \in G~~.$$
When the equations of motion are invariant $ \Phi_a^* H -H \equiv \rho(a)$ is a constant 
on $M$ and $\rho(ab)=\rho (a)+\rho (b)$, $\forall a,b \in G$, so that $a \mapsto \rho(a)$ 
is a homomorphism of $G$ into ${\sf R}$. If $G$ is compact then $\rho (a)=0$, and $H$ is 
invariant to the action of $G$. \\ \indent
The Poisson bracket of the functions $f$, $g$ on $M$ is $ \{ f, g \} = \omega( X_f, X_g)$ 
and 
$$ \frac{d}{dt} ( f \circ F_t) = \{ f \circ F_t, H \} $$
or
$$ \dot{f} = df(X_H)=L_{X_H} f = \omega ( X_f, X_H) = \{ f, H \}~~. $$
\section{Appendix 4: The rigid body and geodesic motion}
Let $\{ x_i, i=1,2,3 \}$ be the space coordinates in the laboratory frame, and 
$\{ x_k' = \sum_i R_{ki} x_i  \}$, the coordinates in the rotated  (intrinsic) frame. 
The rotation matrix $R$ can be taken of the form $R = e^{ \psi J_3} 
e^{ \theta J_1} e^{ \varphi J_3}$, where $( \psi, \theta, \varphi )$ are the
Euler angles, and the ${\sf so}(3,{\sf R})$ generators 
$(J_1, J_2, J_3)$   are $3 \times 3$ matrices with elements $(J_i)_{kl} =
\epsilon_{ikl}$, and commutation relations
$[J_i, J_k] = - \epsilon_{ikl} J_l$. \\ \indent
If   $( \psi, \theta, \varphi )$ depend on time, then 
$$\dot{R} = R \sum_i \omega_i J_i = \sum_i \omega_i' J_i R $$
defines  the angular velocity components $\omega_i$, $\omega_i'$ ($\omega_k'= \sum_i R_{ki} \omega_i$ ) in the 
laboratory, respectively in the intrinsic frame. Explicitly,
$$ \omega_1'= \dot{\theta} \cos \psi + \dot{ \varphi} \sin \theta \sin \psi $$
$$ \omega_2'= - \dot{\theta} \sin \psi + \dot{ \varphi} \sin \theta \cos \psi $$
$$ \omega_3'=   \dot{\psi}  + \dot{ \varphi} \cos \theta ~~. $$
For the rigid body the kinetic energy is $T= \sum_{pq} I_{pq} \omega_p' \omega_q' /2 = \sum_{ij} g_{ij} 
\omega_i \omega_j /2$, where $I_{pq} = I_p \delta_{pq}$ is the intrinsic moment 
of inertia (constant), and $g_{ij} = \sum_{pq} R^{-1}_{ip} R^{-1}_{jq} I_{pq}$. Thus, $L_k = 
\partial T / \partial \omega_k $,  ( $L_k' = \partial T / \partial \omega_k' $), are the angular momentum components, 
and with
$$ \dot{g}_{ij} =  \sum_{k,l=1}^3 \epsilon_{ikl} \omega_k g_{lj} + \epsilon_{jkl} \omega_k g_{li}~~, $$
the conservation law $\dot{L}_k =0$, $k=1,2,3$, yields the Euler equations 
$$ \dot{\bf L'} = {\bf L'} \times {\bf \omega'}~~,$$
or $ g \dot{\bf \omega} = {\bf L } \times {\bf \omega}$. \\ \indent
It is interesting to remark that if $g_{ij}$ is considered as a metric tensor, and 
$$\Gamma^m_{ij} = \sum_{l=1}^3  \frac{ g^{ml}}{2} ( g_{il,j} + g_{jl,i} - g_{ij,l}) $$
as Christoffel symbols of a Riemannian connection, then for a geodesic $\gamma (t)$, $\dot{\gamma} = \eta$, the 
equation $ \dot {\eta}_m = -  \sum_{ij} \Gamma^m_{ij} \eta_i \eta_j$ takes 
the  form $ g \dot{\bf \eta} = 2 (g {\bf \eta }) \times {\bf \eta}$.      
 
\end{document}